\documentclass[aps, prd,reprint,  amsfonts,
  amssymb, amsmath, showpacs, preprintnumbers,letterpaper,superscriptaddress,
  nofootinbib,urlcolor=black,linkcolor=black]{revtex4-2}

\usepackage{babel}
\usepackage{amssymb}
\usepackage{array}

\usepackage{cancel}
\usepackage{braket,bm}
\usepackage[pass]{geometry}
\usepackage{color}
\usepackage{mathrsfs,mathtools}
\usepackage{graphicx}
\usepackage{bbm}
\usepackage{enumerate}
  \DeclareMathAlphabet{\mathpzc}{OT1}{pzc}{m}{it}
\usepackage{subfigure}
\usepackage{graphicx}
\usepackage{float}
\usepackage{soul}
\usepackage{comment}
\usepackage{xtab,afterpage}
\usepackage{lipsum}

\usepackage[colorlinks,citecolor=black]{hyperref}

\def\d{\mathrm{d}}

\def\e{\mathrm{e}}
\def\T{\mathcal{T}}

\usepackage{xcolor}

\definecolor{kblue}{rgb}{0.35, 0.51, 0.8} 

\definecolor{kblued}{rgb}{0.2, 0.396, 0.643} 

\begin{document}



\title{Instability of bubble expansion at zero temperature}

\author{Wen-Yuan Ai} \email{wenyuan.ai@kcl.ac.uk}
\affiliation{Theoretical Particle Physics and Cosmology, King’s College London,\\ Strand, London WC2R 2LS, UK}

\author{Juan S. Cruz} \email{jcr@sdu.dk}
\affiliation{CP3-Origins, Center for Cosmology and Particle Physics Phenomenology, University of Southern Denmark, Campusvej 55, 5230 Odense M, Denmark}

\author{Bj\"{o}rn Garbrecht}
\email{garbrecht@tum.de}

\author{Carlos Tamarit}
\email{carlos.tamarit@tum.de}

\affiliation{Physik Department T70, James-Franck-Stra\ss e,\\ Technische Universit\"{a}t M\"{u}nchen, 85748 Garching, Germany}

\begin{abstract}
In the context of false vacuum decay at zero temperature, it is well known that bubbles expand with uniform proper acceleration.  We show that this uniformly accelerating expansion suffers from an instability related to the bubble size. This can be observed in Minkowski spacetime as a tachyonic mode in the spectrum of fluctuations for the energy functional in the reference frame in which the uniformly accelerating bubble wall appears static. In such a frame, arbitrary small perturbations cause an amplifying departure from the static wall solution. This implies that the nucleated bubble is not a critical point of the energy functional in the rest frame of nucleation but becomes one in the accelerating frame. The aforementioned instability for vacuum bubbles can be related to the well-known instability for the nucleated critical static bubbles during finite-temperature phase transitions in the rest frame of the plasma. It is therefore proposed that zero-temperature vacuum decays as seen from accelerating frames have a dual description in terms of finite-temperature phase transitions.
\end{abstract}

\preprint{KCL-PH-TH/2022-21}
\preprint{TUM-HEP-1415-22}

\maketitle






\section{Introduction}
\label{sec:introduction}

There are many phenomenological scenarios where multiple vacua appear including the Standard Model (SM) of particle physics. Because of the running of the Higgs self-coupling in the SM, the effective Higgs potential develops a lower minimum at a very large field value~\cite{Cabibbo:1979ay, Hung:1979dn, Lindner:1985uk, Sher:1988mj, Sher:1993mf, Degrassi:2012ry, Buttazzo:2013uya}.
Therefore, the electroweak vacuum in the SM is believed to be a false vacuum which can decay to the lower minimum via quantum tunneling.\footnote{In field theory with infinite spatial volume, when the multiple vacua are degenerate tunneling rates between one vacuum and another are vanishing and one has spontaneous symmetry breaking. However, when the spatial volume is finite, tunneling effects become important again \cite{Alexandre:2022qxc, Alexandre:2021imu, Alexandre:2022sho}.} The theory of false vacuum decay is developed in the seminal papers by Coleman and Callan~\cite{Coleman:1977py, Callan:1977pt}, following earlier works~\cite{Langer:1967ax, Langer:1969bc, Kobzarev:1974cp}. When the false vacuum decays in a first order phase transition, i.e. overcoming a classical energy barrier, a bubble nucleates spontaneously and subsequently expands. Vacuum transitions also occur at finite temperature~\cite{Affleck:1980ac, Linde:1980tt, Linde:1981zj}. One notable example was believed to be the electroweak phase transition in the SM. It is now known that it corresponds to a crossover instead~\cite{Kajantie:1996mn,Gurtler:1997hr,Csikor:1998eu}. Nevertheless, a variety of models beyond the SM~\cite{Barger:2007im, Profumo:2007wc, Cline:1996mga, Fromme:2006cm, Dorsch:2013wja, Gunion:1989ci, FileviezPerez:2008bj, Carena:1997ki} do feature first-order phase transitions. 

Bubble nucleation also plays an important role for a variety of phenomena in several contexts. The collisions and mergers of the bubbles produce a potentially observable stochastic background of gravitational waves~\cite{Witten:1984rs, Kosowsky:1991ua, Kosowsky:1992vn, Kamionkowski:1993fg} (see Refs.~\cite{Binetruy:2012ze, Caprini:2015zlo, Cai:2017cbj, Weir:2017wfa, Caprini:2019egz, Hindmarsh:2020hop} for reviews). Moreover, the bubbles of a strong first-order electroweak phase transition may turn out to be pivotal for generating the cosmic matter-antimatter asymmetry~\cite{Kuzmin:1985mm,Shaposhnikov:1987tw}, see also Refs.~\cite{Trodden:1998ym,Morrissey:2012db,Konstandin:2013caa,Garbrecht:2018mrp}. In describing these phenomena, a key parameter is the bubble wall velocity.

The motion of the bubble wall is very different in phase transitions at zero temperature compared to those at finite temperature. In the former, the bubble wall expands with uniform proper acceleration. This standard picture follows from the equation of motion of the scalar field in Minkowski spacetime. As observed by Coleman~\cite{Coleman:1977py}, a solution that is directly tied to the quantum tunneling process and describes an expanding vacuum bubble can be easily obtained by analytical continuation of the $O(4)$-symmetric Coleman bounce, a configuration in Euclidean space (see Sec.~\ref{sec:standard} for a brief review). In this paper we will refer to the vacuum bubble with uniform proper acceleration as a {\it Coleman bubble}. It is widely accepted that the latter describes the real-time growth of nucleated bubbles at zero temperature, and studies of bubble interactions in vacuum transitions typically use unperturbed Coleman bubbles as initial conditions~\cite{Hawking:1982ga,Kosowsky:1991ua,Gould:2021dpm}. Aside from the Coleman $O(4)$ bounce, there also exists a time-independent bounce solution, for which the symmetry is reduced to $O(3)$ both at zero and finite temperature. Only in the latter case is the static bounce connected to the transition rate due to thermal fluctuations. We will refer to the $O(3)$ bounce also as the {\it static bounce}. The static bounce at zero temperature has no direct connection to tunneling processes, but at finite temperature it provides the initial conditions for bubbles nucleated by thermal processes. However, one cannot conclude that the corresponding bubbles remain static after nucleation, as one might be led to think, through a na\"{i}ve analytical continuation of the static bounce. This is because the bubbles nucleated at finite temperature are not stable even when they are critical points of the free energy. Shortly after the nucleation of a bubble at finite temperature, before the plasma backreaction is sizable, the critical finite-temperature bubble will undergo a period of acceleration triggered by some perturbations. For studies of bubble dynamics during thermal transitions, the initial condition for bubble configurations is usually different from the exact static critical bubble, see e.g. \cite{Hindmarsh:2015qta, Hindmarsh:2017gnf}, as otherwise the bubbles would not expand. This is in contrast to the case of vacuum transitions, for which as mentioned above one usually evolves unperturbed Coleman bubbles.

In this article, the uniformly accelerating motion of the bubble wall at {\it zero temperature} is revisited. We show that although the standard acceleration satisfies the equation of motion, it is not stable under small perturbations of the bubble radius, which preserve the spherical symmetry of the bubble. This is seen in the uniformly accelerating frame where the expanding bubble appears to be static. We derive the eigenvalue equation for the fluctuations of the energy functional in the static background of the scalar field and show that there is a tachyonic mode. 

The Coleman bubble appears to be static in an accelerating frame and suffers from an instability akin to the case of the static critical bubble in finite-temperature phase transitions in the \emph{plasma frame}, i.e. the rest frame of the plasma. Indeed, there is also a tachyonic mode in the background of the static critical bubble in finite-temperature phase transitions. This suggests that false vacuum decay at zero temperature, observed in an accelerating frame, can be viewed as a thermal transition. This correspondence was pointed out for bubbles nucleating around horizons in Ref.~\cite{Ai:2018rnh} and has been used implicitly in many studies~\cite{Mukaida:2017bgd,Shkerin:2021zbf,Shkerin:2021rhy,Strumia:2022jil,Briaud:2022few}. For uniformly accelerating observers, the presence of the Rindler horizon implies the perception of a ``plasma''---the Unruh bath~\cite{Unruh:1976db}. A natural consequence of the instability of the Coleman bubble under radial perturbations, as well as the correspondence with finite-temperature phase transitions, is that for classical simulations of the expansion of bubbles in vacuum transitions one may consider bubble configurations that are different from the exact Coleman critical bubble, in analogy with the usual initial conditions for simulations of thermal transitions. In principle, to study the late-time growth of vacuum bubbles in the accelerating frame or of finite-temperature bubbles in the plasma frame, one has to consider the coupled system between the scalar field and the Unruh bath or the plasma~\cite{Steinhardt:1981ct, Liu:1992tn, Ignatius:1993qn, Moore:1995ua, Moore:1995si, Konstandin:2014zta, Ai:2021kak}, respectively. It is expected in either case that the walls of the bubbles will, in general, travel with a nonvanishing velocity in the aforementioned frames.

Another conceptual point about the quantum nucleation of classical bubbles at zero temperature that needs clarification concerns Lorentz invariance. The full analytic continuation of the Euclidean bubbles are hyperbolic three-spaces, or three-dimensional hyperboloids, and there are infinitely many spatial hypersurfaces---related by boosts---that are normal to the four-velocity of the bubble wall. In order to interpret the analytically continued configurations in terms of real-time processes, the full three-dimensional hyperboloid may be truncated by a particular flat spatial hypersurface. Setting the time to be constant on this hypersurface defines a particular inertial frame and we will refer to it as the {\it rest frame of nucleation}. The first related question is: What selects the rest frame for a nucleation event? This issue has become a subject of investigation~\cite{Garriga:2012qp,Garriga:2013pga,Chen:2020lrl}. Another question is, for a bubble observed at a {\it late} time, how one can identify the rest frame of nucleation by performing measurements only on the late-time bubble wall motion.\footnote{One may think that the rest frame of nucleation is simply the rest frame of the bubble center and the latter is unique. This is not correct because there is no unique criterion to tell where is the bubble center. The $SO(3,1)$ symmetry of the full three-dimensional hyperboloid allows any inertial observer to identify a static bubble center at sufficiently late times.} The instability of the uniformly accelerating expansion mentioned above can in principle provide a way to determine the rest frame of nucleation if one assumes that perturbations on the bubble wall occur at the time of bubble nucleation so that the unstable behavior starts also at that time. Analogously, in the case of finite-temperature phase transitions the instability can also be used to determine the rest frame of the plasma {\it solely} from the bubble wall motion.

The present paper is organized as follows. We review false vacuum decay at zero and finite temperature and the standard bubble growth for zero-temperature bubbles in Sec.~\ref{sec:standard}. In Sec.~\ref{sec:instability}, we focus on zero temperature transitions and introduce the uniformly accelerating frame, where the standard bubble motion appears to be static. We then analyze the fluctuations around the static background and prove the existence of a tachyonic mode in the fluctuation spectrum.  An analogous instability is shown to exist for the bubbles nucleated at finite temperature as critical points of the free energy in the plasma frame. In Sec.~\ref{sec:correspondence}, we discuss the correspondence between the false vacuum decay at zero temperature as seen from the uniformly accelerating observers and finite-temperature phase transitions. This provides a physical picture of the instability for the standard accelerating expansion of zero-temperature bubbles. We conclude in Sec.~\ref{sec:conclusion}. Throughout this paper, we use $\hbar=c=1$ and the metric signature  $(+,-,-,-)$ in Minkowski spacetime.

\section{Bubble aspects in phase transitions}
\label{sec:standard}

\subsection{At zero temperature}

In this section, we briefly review the Callan-Coleman formalism~\cite{Callan:1977pt} of false vacuum decay and the uniformly accelerating growth of  nucleated bubbles~\cite{Coleman:1977py}. Additionally, we summarize the main features of bubble nucleation in finite-temperature phase transitions.
\begin{figure}[htb]
  \centering
  \includegraphics[scale=0.5]{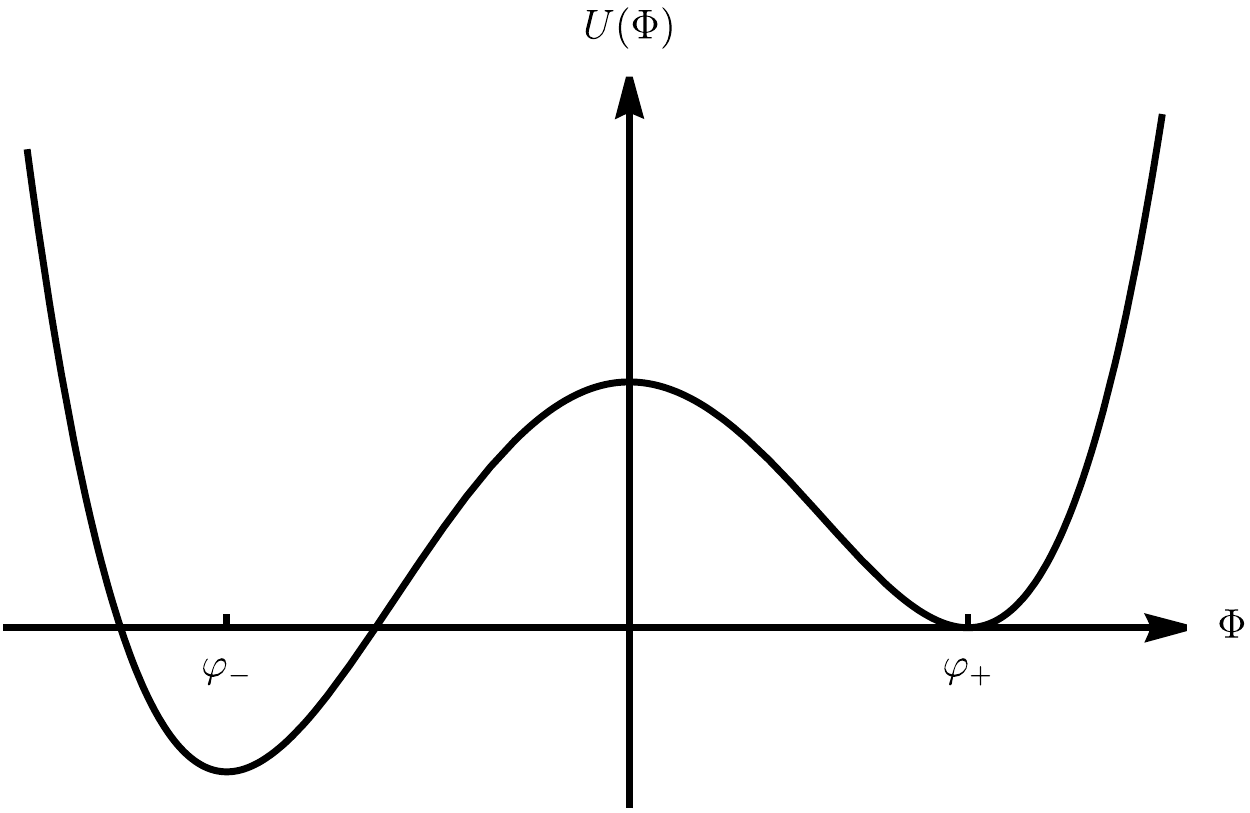}
  \caption{A classical potential $U(\Phi)$ that possesses a false vacuum, $\varphi_+$, and a true vacuum, $\varphi_-$. The potential is chosen such that $U(\varphi_+)=0$ for convenience. \label{fig:potential}}
\end{figure}

In a scalar field theory with a potential of the form shown in Fig.~\ref{fig:potential}, the ground state about the metastable minimum $\varphi_+$ is not stable and can decay through quantum tunneling. In order to obtain the decay rate, Callan and Coleman consider the following Euclidean false vacuum to false vacuum transition amplitude\footnote{The calculation of the rate of quantum tunneling directly in Minkowski spacetime has been the subject of some recent research~\cite{Cherman:2014sba,Tanizaki:2014xba,Ai:2019fri,Matsui:2021oio,Hayashi:2021kro}.}
\begin{align}
\label{partition}
Z[0] = \langle \varphi_+| \e^{-H\T}|\varphi_+\rangle = \int\mathcal{D}\Phi\:\e^{-S_{\rm E}[\Phi]},
\end{align}
where $H$ is the full Hamiltonian and $\T$ is the amount of the Euclidean time for this transition amplitude. The classical Euclidean action is given by
\begin{align}
\label{E-action}
S_{\rm E}[\Phi]=\int \d^4x\; \left[\frac{1}{2}\delta^{\mu\nu}(\partial_\mu\Phi)\partial_\nu\Phi+U(\Phi)\right],
\end{align}
where $\delta^{\mu\nu}$ is the Kronecker symbol and $\mu,\nu=1,...,4$. The potential is chosen for convenience such that $U(\varphi_+)=0$ but otherwise kept general.

To see why the Euclidean partition function gives out the decay rate, we insert a complete set of energy eigenstates into the partition function, i.e.,
\begin{align}
\label{partition2}
\langle\varphi_+|\e^{-H\T}|\varphi_+\rangle = \sum\limits_n\,\e^{-E_n \T}\:\langle\varphi_+|n\rangle\langle n|\varphi_+\rangle.
\end{align}
Taking the large $\T$ limit, we thus obtain
\begin{align}
E_0=-\lim_{\T\rightarrow\infty}\frac{1}{\T}\ln\left(\frac{Z[0]}{|\langle\varphi_+|0\rangle|^2}\right).
\end{align}
In Ref.~\cite{Callan:1977pt}, it is shown from the path integral expression for the partition function that the energy $E_0$ has an imaginary part which gives the decay rate as
\begin{align}
\varGamma=-2\;{\rm Im} E_0=\lim_{\T\rightarrow\infty}\frac{2}{\T}{\rm Im}(\ln Z[0]).
\end{align}
Here we have used the fact that the squared amplitude does not contribute to the imaginary part.

One can calculate the path integral by expanding it around the stationary points. The dominant contributions to the tunneling rate are due to the lowest-lying \emph{bounce}
solution (and the fluctuations about it).
The (tree-level) bounce is a solution to the classical equation of motion
\begin{align}
\label{eommmmmmm}
- \partial^2\varphi + U'(\varphi) = 0
\end{align}
that satisfies the boundary conditions
$\varphi|_{x_4\rightarrow\pm\infty}=\varphi_+$ and $\dot{\varphi}|_{x_4=0}=0$, where the dot denotes the derivative with respect to $x_4\equiv\tau$ and the prime denotes the derivative of the classical potential  with respect to the field $\varphi$.
The bounce solution has an $O(4)$ symmetry and thus is a function only of $\rho=\sqrt{{\bf x}^2+\tau^2}\equiv \sqrt{r^2+\tau^2}$. Equation~\eqref{eommmmmmm} then reduces to
\begin{align}
\label{Eq7}
-\frac{\d^2}{\d\rho^2}\varphi-\frac{3}{\rho}\frac{\d}{\d\rho}\varphi+U'(\varphi)=0.
\end{align}
We denote the bounce solution as $\varphi_{b}$.
The decay rate per unit volume is given as
\begin{align}
\varGamma/V=A\; \e^{-B},
\end{align}
where
\begin{align}
\label{bounce-action}
B\equiv 2\pi^2\int\d \rho\,\rho^3\left[\frac{1}{2}(\partial_\rho\varphi_b)^2+U(\varphi_b)\right]
\end{align}
is the bounce action and $A$ is a prefactor that can be obtained from the fluctuations about the bounce.

From Eqs.~\eqref{partition} and~\eqref{E-action}, the fluctuations can be studied through the eigenvalue equation
\begin{align}
\label{eigenEq}
\left[-\partial^2+U''(\varphi_b)\right]\hat{\Phi}_n(x)=\lambda_n \hat{\Phi}_n(x).
\end{align}
Because of the $O(4)$ symmetry of the bounce, one can separate the angular dependence as
\begin{align}\label{eq:hyperspherical}
\hat{\Phi}_n(x)=\phi_{nj}(\rho)Y_{jlm}({\bf e}_r),
\end{align}
where $Y_{jlm}({\bf e}_r)$ are hyperspherical harmonics evaluated on a vector ${\bf e}_r$ on the unit three-sphere. Substituting this decomposition into Eq.~\eqref{eigenEq}, one obtains the radial eigenvalue equation
\begin{align}
\label{radial-eigen}
\left[-\frac{\d^2}{\d \rho^2}-\frac{3}{\rho}\frac{\d}{\d \rho}+\frac{j(j+2)}{\rho^2}+U''(\varphi_b(\rho))\right]\phi_{nj}(\rho)=\lambda_n\phi_{nj}(\rho),
\end{align}
where we impose the boundary conditions $\phi_{nj}(\infty)=0$.
It is well known that there is a negative mode with $j=0$~\cite{Callan:1977pt,Garbrecht:2018rqx} which is responsible for the imaginary part in the partition function. For instance, for the thin-wall limit which applies when the energy difference between the false and true vacua is much smaller than the barrier height, we have~\cite{Garbrecht:2015oea}
\begin{align}
\label{Fluct}
\left[-\frac{\d^2}{\d \rho^2}-\frac{3}{\rho}\frac{\d}{\d \rho}+U''(\varphi_b(\rho))\right]\partial_\rho\varphi_b(\rho)=-\frac{3}{R_c^2}\partial_\rho\varphi_b(\rho),
\end{align}
where $R_c$ is the bounce radius.

It is the case that in the thin-wall limit the negative mode corresponds to dilatations of the bounce solution. In this regime, one can separate the bounce action into two parts, the surface contribution and the volume contribution. The surface term is
\begin{align}
\label{eq:Bsurface}
B_{\rm surface}(R)&=2\pi^2\int_{R-\delta/2}^{R+\delta/2}\d \rho\,\rho^3\left[\frac{1}{2}(\partial_\rho\varphi_b)^2+U(\varphi_b)\right]\notag\\
&=2\pi^2 R^3\sigma,
\end{align}
where we have defined the surface tension $\sigma$ and $\delta$ is a small number representing the thickness of the bubble wall. The volume term is
\begin{align}
\label{eq:Bvolume}
B_{\rm volume}(R)=2\pi^2\int_{0}^{R}\d \rho\,\rho^3 U(\varphi_-)=-\frac{\pi^2}{2}R^4\epsilon,
\end{align}
where $\epsilon\equiv -U(\varphi_-)$. Then $R_c$ is determined by extremizing the bounce action with respect to $R$,
\begin{align}
\label{bounce-action2}
\left.\frac{\d B}{\d R}\right|_{R=R_c}=0&=6\pi^2 R_c^2\sigma-2\pi^2 R_c^3\epsilon\notag\\
&=\frac{3\pi}{2}\left(4\pi R_c^2\sigma-\frac{4}{3}\pi R_c^3\epsilon\right),
\end{align}
giving $R_c=3\sigma/\epsilon$. Substituting $R_c$ into $B(R)$, one obtains the standard result $B(R_c)=27\pi^2\sigma^4/(2\epsilon^3)$. The negative eigenvalue is given by~\cite{Garbrecht:2015oea}\footnote{The expression given in Ref.~\cite{Garbrecht:2015oea} misses the factor $1/4$ because the normalized dilatational mode in the thin-wall limit is given by $2 B^{-1/2}\partial_{R_c}\varphi$. The normalization is relevant in order to quantitatively relate variations by the negative modes to dilatations. This factor can be checked from Eqs.~\eqref{eq:Bsurface} and~\eqref{eq:Bvolume} with $B(R)=B_{\rm surface}(R)+B_{\rm volume}(R)$.}
\begin{align}
\label{negative-eigenvalue}
\lambda_0=-\frac{3}{R_c^2}=\frac{1}{4}\left.\frac{1}{B(R)}\frac{\d^2B(R)}{\d R^2}\right|_{R=R_c},
\end{align}
indicating that the negative mode corresponds to dilatations of the bounce.

The existence of a negative mode in the eigenvalue equation~\eqref{radial-eigen} is a characteristic property of a potential allowing for false vacuum decay, regardless of the thin-wall approximation. This fact will turn out to be important when we discuss the fluctuations about the bubble, which appears static in the comoving frame that is accelerating with the wall.

In the rest frame of nucleation, the bounce solution yields the nucleated field configuration via $\varphi_{\rm bubble}(t=0,{\bf x})=\varphi_b(\tau=0,{\bf x})$. In fact, in a tunneling process, there should be a wave functional for the nucleated scalar configurations and $\varphi_b(\tau=0,{\bf x})$ should only be the most probable one. In this paper, we assume that the nucleated configuration is given by, or close enough to, $\varphi_b(\tau=0,{\bf x})$. In the thin-wall case, $\varphi_b(\tau=0,{\bf x})$ is a bubble of radius $R_c$.

\begin{figure}[htbp]
  \centering
  \includegraphics[scale=0.5]{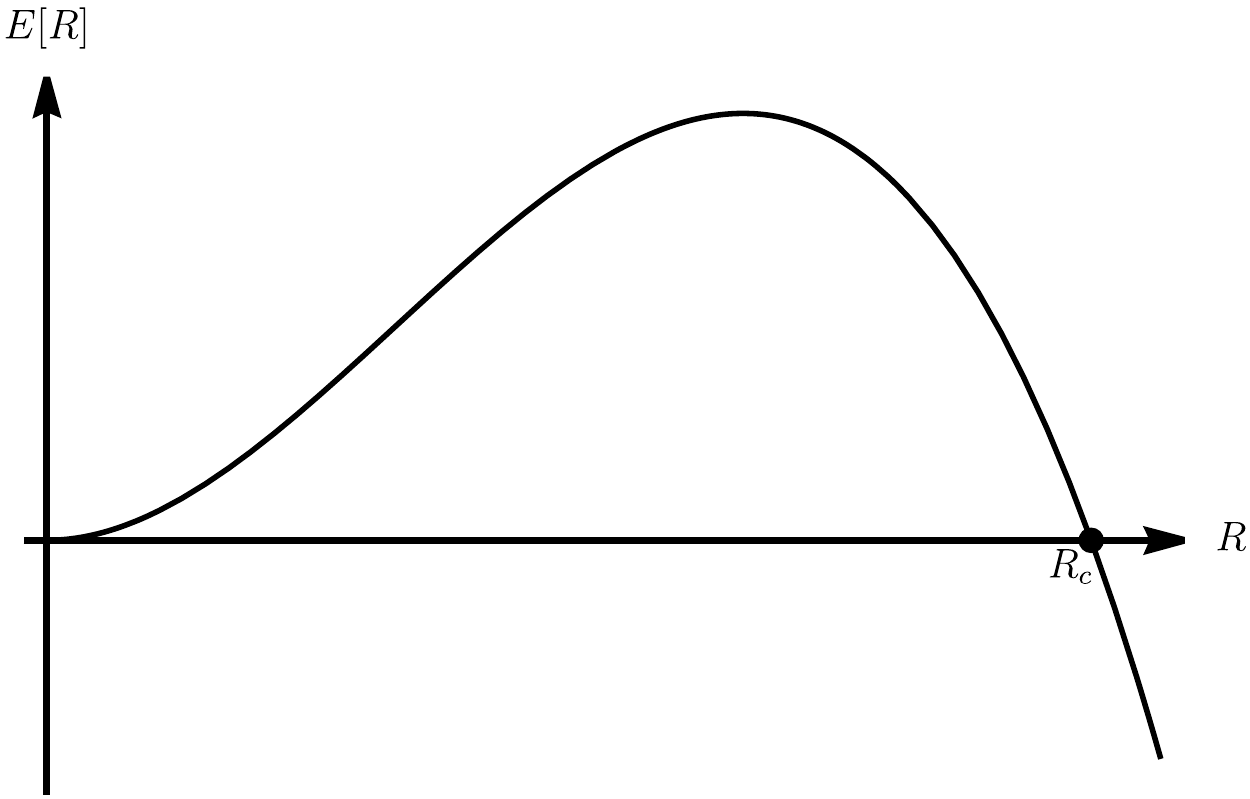}
  \caption{Dependence of the bubble energy on the bubble radius, at $t=0$ in the rest frame of nucleation and the thin-wall approximation. The radius of the nucleated bubble corresponds to an energy-conserving point, indicated by the dot. \label{fig:ER}}
\end{figure}

We want to remark that $R_c$ corresponds to a local maximum of of the action, $B(R)$, but not of the energy. Indeed, in the thin-wall case, the energy for a static bubble of radius $R$ is given by $E(R)=4\pi R^2\sigma-4\pi R^3\epsilon/3$. From Eq.~\eqref{bounce-action2} one sees that $E(R_c)=0$ which is reasonable because quantum tunneling satisfies energy conservation. A consequence of this is that in the rest frame of nucleation, the nucleated bubble at zero temperature is {\it not} critical according to the criterion of successful nucleation. This is because it will still expand if its radius is slightly smaller than $R_c$ (see Fig.~\ref{fig:ER}). Yet, the bubble can be viewed as critical in the sense that it is the configuration most likely to occur. This is further confirmed in the numerical calculations in Sec.~\ref{sec:instability}. For clarification of terminology, we define
\begin{itemize}
\item {\it critical or H-critical:} The bubble is called critical or H-critical if it is a saddle point of the energy (or free energy for the finite temperature case) functional.
\item {\it S-critical:} The bubble is called S-critical if it is a saddle point of the four-dimensional Euclidean action.
\end{itemize}
With the above terminology, in the rest frame of nucleation the nucleated bubble at zero temperature is S-critical but not H-critical. As we shall see shortly, the nucleated bubble is however H-critical in the accelerating frame, just as nucleated bubbles in finite-temperature phase transitions are H-critical in the plasma frame.

\begin{figure}[htbp]
  \centering
  \includegraphics[scale=0.4]{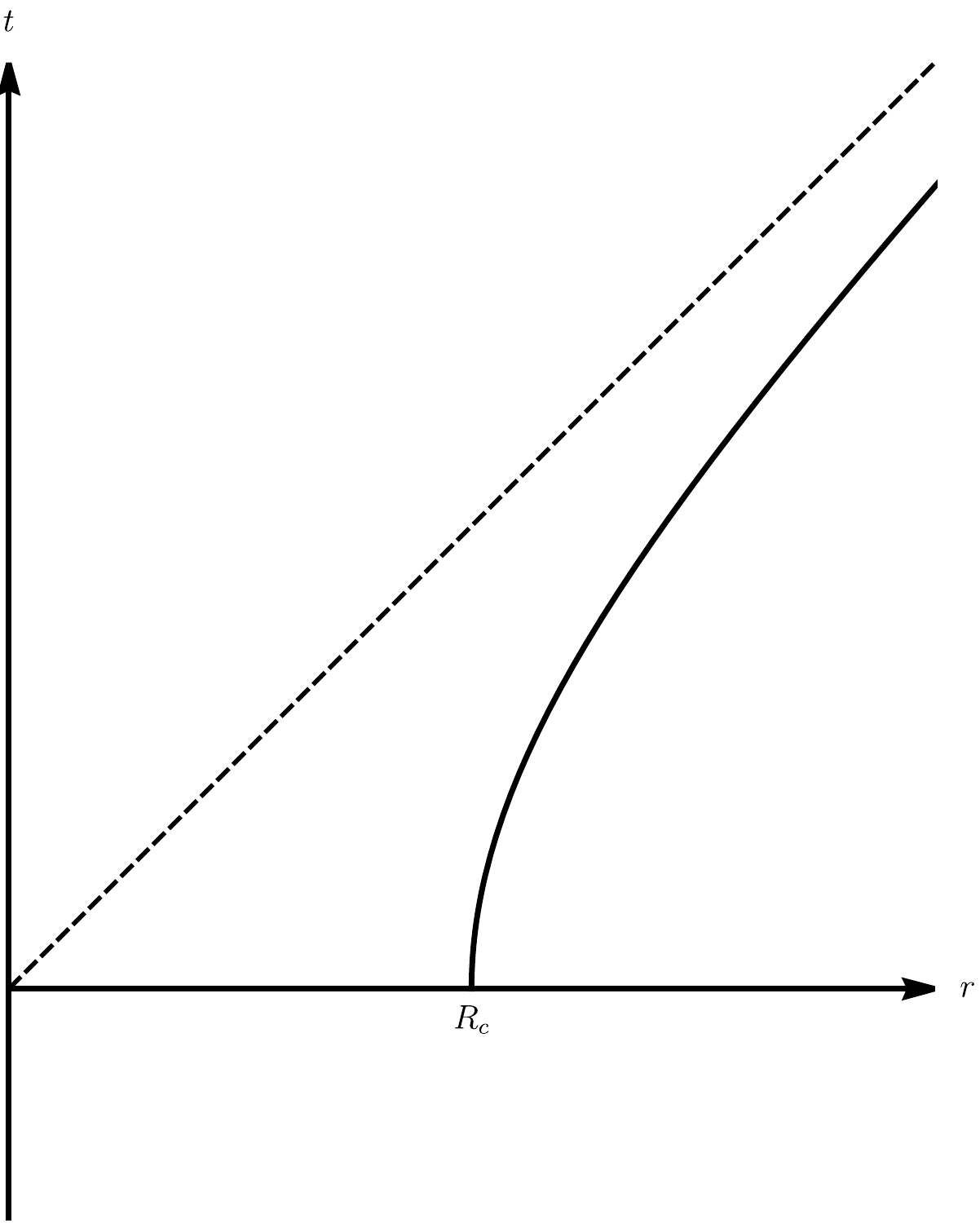}
  \caption{The uniformly accelerated motion of the bubble wall after the nucleation at $t=0$. The solid hyperbola represents the bubble wall while the dashed line represents the light cone. \label{fig:standard}}
\end{figure}

The subsequent evolution after nucleation follows from the equation of motion in Minkowski spacetime. Since it can be obtained from the Euclidean one by an inverse Wick rotation $\tau\rightarrow {\rm i}t$, it directly follows that when the nucleated bubble is given by $\varphi_b(\tau=0,{\rm x})$, the bubble would evolve as $\varphi_{\rm bubble}(t,{\bf x})=\varphi_b({\rm i}t,{\bf x})=\varphi_b(\sqrt{-t^2+r^2})$. For definiteness, take the position of the bubble wall as the set of spacetime points corresponding to some fixed field value between $\varphi_+$ and $\varphi_-$. Since the value of the field $\varphi_{\rm bubble}(t,{\bf x})$ is constant on the surface $r^2-t^2={\rm const}$, the worldline of the bubble wall in a fixed spatial direction ${\bf e}_{\bf x}$ is thus a hyperbola, as shown in Fig.~\ref{fig:standard}. It is therefore concluded in Ref.~\cite{Coleman:1977py} that bubbles nucleated at zero temperature grow at a uniform proper acceleration rate. Note that this standard picture of bubble nucleation and subsequent growth is only true for the rest frame of nucleation. For the same spacetime diagram (the truncated three-dimensional hyperboloid), a boosted observer would see a different bubble-growth history.
Yet the wordlines of the bubble are then still given by hyperbolae. Thus, the rest frame of nucleation is an inertial frame where an S-critical bubble is nucleated with vanishing bubble wall velocity at $t=0$ and subsequently expands at uniform acceleration---when there are no perturbations about the bubble configuration.

\subsection{At finite temperature}

In addition to vacuum transitions and the Coleman bubble, finite-temperature phase transitions are also of significant interest. The relevant thermal partition function at a high temperature $T=1/\beta$ is
\begin{align}
\label{partitionFiniteT}
Z[T] =\e^{-\beta F(T)}= {\rm Tr}\, \e^{-\beta H} = \int\mathcal{D}\Phi\:\e^{- S_{\rm E}[\Phi]}.
\end{align}
In the equation above, $F$ is the total free energy, while $S_{\rm E}$ is the Euclidean action with the compact time interval $\beta$ and periodic boundary conditions. The relevant saddle point solutions can be found from reducing Eq.~\eqref{E-action} to time-independent configurations,\footnote{At finite temperature, one can still consider the $O(4)$ Coleman bounce solution provided its radius is smaller than the inverse temperature. However, at high temperature the corresponding bounce action should be larger than that of the $O(3)$ static bounce, and thus nucleation via the Coleman bounce is suppressed. In this paper, we only consider the static bounce at finite temperature.}
\begin{align}
\label{E-action2}
S_{\rm E}[\Phi]/\beta \rightarrow S_{3}[\Phi]=\int \d^3x\; \left[\frac{1}{2}\,(\partial_i \Phi) \partial_i\Phi+U(\Phi)\right].
\end{align}
In analogy to the vacuum case, the transition rate induced by thermal effects is
\cite{Linde:1980tt}
\begin{align}
 \varGamma= -2 {\rm Im} F(T) \sim \exp(-\beta B_3)\,,
\end{align}
with $\sim$ indicating that the relation holds up to the determinant prefactor. Above, $B_3=S_3[\bar\varphi_b]$ designates the three-dimensional action evaluated on a static bounce solution $\bar\varphi_b$ with the $O(3)$ symmetry, which is an extremum of $S_3$ and thus satisfies
\begin{align}
\label{eq:staticbubble}
-\frac{\d^2}{\d r^2}\,\bar\varphi_b-\frac{2}{r}\frac{\d}{\d r}\bar\varphi_b+U'(\bar\varphi_b)=0,
\end{align}
with
\begin{align}
\bar\varphi_b(\infty)=\varphi_+,\quad \bar\varphi'_b(0)=0.
\end{align}
Since thermal fluctuations can induce significant corrections to the classical potential, it is customary to include the leading effects of thermal fluctuations in solving the bounce. In such a case, one would replace the classical potential $U(\Phi)$ with one including thermal corrections in $S_3[\Phi]$. $S_3[\varphi]$ can then be thought as the free energy in a scalar background field $\varphi$. Below, we will use the same notation, but $U(\Phi)$ could include the leading finite-temperature corrections.\footnote{For false vacuum decay at zero temperature, one can also compute the bounce with a quantum-corrected potential.}  For the purpose of the computation of the static bounce, the temperature can be considered constant and absorbed into the various couplings. The bounce corresponds to an unstable configuration at the top of the free-energy barrier (where the free-energy here includes spatial gradients) between the true and the false vacuum. Because of this instability, one naturally expects a negative mode in the fluctuation operator $\delta^2 S_3/\delta \varphi^2$ evaluated at the static  bounce.
In analogy with Eq.~\eqref{eq:hyperspherical}, for the three-dimensional action $S_3$ the fluctuations can be decomposed as
\begin{align}
 \hat{\Phi}_n({\bf x})=\phi_{nl}(r) Y_{lm}(\bf e_{x}),
\end{align}
where $\bf x$ denotes the three-dimensional spatial coordinates, and $Y_{lm}(\bf e_x)$ are the usual three-dimensional spherical harmonics evaluated on the vector $\bf e_x$ on the unit {\it two}-sphere. Analogously to Eq.~\eqref{radial-eigen}, one obtains now a radial eigenvalue equation
\begin{align}
\label{radial-eigen-finite-T}
\left[-\frac{\d^2}{\d r^2}-\frac{2}{r}\frac{\d}{\d r}+\frac{l(l+1)}{r^2}+U''(\bar{\varphi}_b(r))\right]\phi_{nl}(r)=\lambda_n\phi_{nl}(r).
\end{align}
By taking derivatives with respect to $r$ in Eq.~\eqref{eq:staticbubble}, it is easily shown that in the thin-wall limit there is a negative eigenvalue in the $l=0$ sector given by
\begin{align}
\label{eq:lambdacstatic}
\lambda_{0,\rm static}=-\frac{2}{R_{c,\rm static}^2}=\frac{1}{3B_3(R_{c,\rm static})}\,\left.\frac{\d^2B_3(R)}{\d R^2}\right|_{R=R_{c,\rm static}},                                                                                                                                                                                                                                                                                                                            \end{align}
where $R_{c,\rm static}$ is obtained by extremizing the three-dimensional action $S_3$ similarly to the zero temperature case.

In contrast to the Coleman bounce, whose analytic continuation to real time gives a uniformly accelerating expansion, the static bounce corresponds to a time-independent solution. It is usually assumed that the scalar configuration at the time of nucleation in finite-temperature phase transitions is very close to the static bounce, but it cannot be restricted to the bounce itself because in that case the bubbles would not expand and the transition would not complete.  Again, one may think that the static bounce, which we will also refer to as the static bubble, only gives the most probable nucleated state. Hence, in this paper we assume that the nucleated bubbles are given by small deformations of the static bounce. In contrast to the Coleman bubble, the static bubble is  H-critical, as $S_3[\varphi]$ is now the free energy. Therefore, it is expected that some deformations of the static bounce will result in an accelerating expansion or contraction at early times, depending on the competition between the surface tension and the bulk free-energy difference between the phases outside and inside the bubble. These expanding (contracting) deformations of the static bounce lead to a successful (failed) phase transition.

In the following section we show that the Coleman bubble also features an instability in its time evolution, in analogy to the static bubble. However, this instability can only be seen in the accelerating frame where the Coleman bubble becomes H-critical. For both the Coleman bubble and the static bubble, the instability can be connected to the negative mode of the fluctuation operator about the corresponding background.

\section{Classical Instability of the Bubble}
\label{sec:instability}

\subsection{Coleman bubble in the accelerating frame}

We show in the present section that the standard bubble growth for zero-temperature transitions reviewed in the previous section is unstable against arbitrarily small perturbations. We first recall some simple facts about the uniform accelerating frame. Consider a worldline $\{t,r,\theta_0={\rm const},\phi_0={\rm const}\}$ satisfying
\begin{align}
\label{eq:hyperbola}
r^2-t^2=\rho^2={\rm const}.
\end{align}
In this coordinates the metric reads $\d s^2=\d t^2-\d r^2-r^2(\d\theta^2+\sin^2\theta\d \phi^2)$.
Taking $\rho=R_c$ would give the worldline of the Coleman bubble for fixed angles $\theta$ and $\phi$. Here we have assumed that the bubble center is located at the origin. From the above equation we immediately obtain for the inertial observer that
\begin{equation}
	\frac{\d^2 r}{\d t^2} = \frac{\rho^2}{(t^2 + \rho^2)^{3/2}},
\end{equation}
so that the coordinate acceleration is not uniform. On the other hand, we can compute the proper acceleration on the worldline by first expressing it through the proper time $\xi$ and then simply taking derivatives. We first parametrize the worldline as
\begin{equation}
\label{eq:gamma_xi}
	\gamma(\xi) = (\rho \sinh( \alpha \xi) , \rho \cosh( \alpha \xi ) , \theta_0, \phi_0)
\end{equation}
so that Eq.~\eqref{eq:hyperbola} is automatically satisfied. To make the argument in the hyperbolic functions dimensionless, we have introduced a variable $\alpha$ which has the same dimension as length and time. It is determined by imposing that the Minkowski norm of the four-velocity of the worldline equals one,  $ g_{\mu\nu}(\d{\gamma^\mu}/\d \xi) (\d \gamma^\nu/\d\xi) = 1$  (so that $\xi$ can indeed be interpreted as the proper time) which gives $\alpha=1/\rho$. Taking the derivative of $\gamma(\xi)$, one obtains the four-velocity,
\begin{equation}
	u(\xi) = (\cosh (\xi/\rho), \sinh(\xi/\rho), 0, 0).
\end{equation}
The proper acceleration is indeed uniform,
\begin{equation}
	g_{\mu\nu}\left(\frac{\d u^\mu}{\d \xi}\right)\left(\frac{\d u^\nu}{\d\xi}\right) = -\frac{1}{\rho^2}.
\end{equation}
Equation~\eqref{eq:gamma_xi} also indicates the usual coordinates in the uniformly accelerating frame $\{\xi,\rho,\theta,\phi\}$.

In order to explicitly show the instability, and inspired by the worldline followed by the wall, we choose the following coordinates $\{t_a,r_a,\theta,\phi\}$ via
\begin{subequations}
\label{frame:acc}
\begin{align}
t&=r_a \sinh t_a,\\
r&=r_a\cosh t_a.
\end{align}
\end{subequations}
Note that $t_a$ is dimensionless and is related to the proper time via $t_a=\xi/r_a$. Still, $r^2-t^2=r_a^2$, meaning that the Coleman bubble appears static in this coordinate frame $\varphi_{\rm bubble}(t_a,r_a,\theta,\phi)=\varphi_{\rm bubble}(r_a)=\varphi_b(r_a)$.  For these new coordinates, we have the metric
\begin{align}
\d s^2=r_a^2\d t_a^2-\d r_a^2-(r_a^2\cosh^2 t_a)(\d \theta^2+\sin^2\theta\, \d\phi^2).
\end{align}

Substituting the above metric into the Minkowskian action
\begin{align}
S_{\rm M}[\Phi]=\int\d^4 x\; \sqrt{-g}\left[\frac{1}{2} g^{\mu\nu}(\partial_\mu\Phi)\partial_\nu\Phi-U(\Phi)\right]
\end{align}
we obtain
\begin{align}
\label{action-co}
S_{\rm M}[\Phi]=\int\d t_a\int\d r_a\int \d\theta\int \d \phi\;\mathcal{L}^a_{\rm M},
\end{align}
where
\begin{align}
\mathcal{L}^a_{\rm M} &=\frac{1}{2}r_a\sin\theta(\cosh^2 t_a)\left(\partial_{t_a}\Phi\right)^2\notag\\
&-\frac{1}{2}r_a^3\sin\theta(\cosh^2 t_a) \left(\partial_{r_a}\Phi\right)^2
-\frac{1}{2}r_a\sin\theta\left(\partial_{\theta}\Phi\right)^2\notag\\
&-\frac{1}{2}r_a\frac{1}{\sin\theta}\left(\partial_{\phi}\Phi\right)^2-r_a^3\sin\theta(\cosh^2t_a) U(\Phi).
\end{align}
Through a Legendre transformation, we obtain the Hamiltonian density,
\begin{align}
	\begin{aligned}
	\label{Hamiltonian}
	{\cal H}[\Pi,\Phi]&=\int\d r_a\int \d\theta\int \d \phi\;\bigg[\frac{1}{2}r_a^2\Pi^2 \\
	&+\frac{1}{2}r_a^3\sin\theta(\cosh^2 t_a) \left(\partial_{r_a}\Phi\right)^2
	+\frac{1}{2}r_a\sin\theta\left(\partial_{\theta}\Phi\right)^2\\
	&\left.+\frac{1}{2}r_a\frac{1}{\sin\theta}\left(\partial_{\phi}\Phi\right)^2+r_a^3\sin\theta(\cosh^2t_a) U(\Phi)\right],
	\end{aligned}
\end{align}
where $\Pi\equiv \partial{\cal L}/\partial_{t_a}\phi$. Note that since $t_a$ is not the proper time, the above Hamiltonian density does not give the physical energy density measured by the uniformly accelerating observers. Nonetheless, such a coordinate frame can still be used for computations and is employed in discussions on the Unruh effect~\cite{Susskind:2005js}. An advantage of using the dimensionless time $t_a$ is that it is  associated with a constant global dimensionless ``temperature'' $\hat T_a=1/(2\pi)$, while the proper temperature in the accelerated frame, $T_a=1/(2\pi r_a)$, is ``inhomogeneous'' in $r_a$. A constant $\hat T_a$ makes it possible to identify the zero-temperature phase transition as a finite-temperature phase transition in the coordinate frame $\{t_a,r_a,\theta,\phi\}$.

Let us consider spherically symmetric and $t_a$-independent field configurations $\Phi(r_a)$. One can then integrate over $\theta$ and $\phi$ in Eq.~\eqref{Hamiltonian} and obtain
\begin{align}
H[\Phi(r_a),t_a]=4\pi (\cosh^2t_a)\int\d r_a\, r_a^3 \left[\frac{1}{2}(\partial_{r_a}\Phi)^2+U(\Phi)\right].
\end{align}
Then $0= \delta H/\delta\phi$ gives
\begin{align}
\label{Eq25}
\frac{\d^2}{\d r_a^2}\Phi+\frac{3}{r_a}\frac{\d}{\d r_a}\Phi-U'(\Phi)=0.
\end{align}
This is exactly the equation of motion for the bounce with $\rho$ replaced by $r_a$, see Eq.~\eqref{Eq7}. Therefore the configuration of a uniformly accelerating bubble, $\varphi_{\rm bubble}(r_a)=\varphi_b(r_a)$, indeed satisfies the Hamilton equations. Furthermore, this means that in the thin-wall case $R_c$ corresponds to an extremum of the energy $H[\varphi(R)]$ in the accelerating frame.

In order to study the stability properties of this configuration, we substitute $\Phi(r_a)=\varphi_b(r_a)+\hat{\Phi}(r_a)$ into Eq.~\eqref{Hamiltonian} and obtain the following eigenvalue equation
\begin{align}
\label{20}
\left[-\frac{\d^2}{\d r_a^2}-\frac{3}{r_a}\frac{\d}{\d r_a}+U''(\varphi_b)\right]\hat{\Phi}_n(r_a)=\lambda_n\hat{\Phi}_n(r_a).
\end{align}
We impose the boundary condition for the fluctuations $\hat{\Phi}(r_a=\infty)=0$. We also require the derivative of $\hat{\Phi}_n$ with respect to $r_a$ to vanish at at $r_a=0$, in order to have a regular behavior. Equation~\eqref{20}, with the replacement $r_a\rightarrow \rho$, is exactly the same as Eq.~\eqref{Fluct} which corresponds to $j=0$ in Eq.~\eqref{radial-eigen}. Thus, we immediately know that there exists a tachyonic mode in the eigenspectrum. We therefore conclude that {\it the static background $\varphi_b(r_a)$ is energetically unstable}; the extreme point is not a minimum.

One can also witness the instability by considering the motion of the bubble wall in the presence of a small perturbation. For this purpose, we allow $\Phi$ to have a dependence on $t_a$, but we keep an $O(3)$ symmetry for the bubble.
Integrating out $\theta$ and $\phi$ in Eq.~\eqref{action-co}, we obtain
\begin{align}
\label{action-co2}
&S_{\rm M}[\Phi]=4\pi\int\d t_a\d r_a\,\notag\\
&\times(\cosh^2 t_a)r_a^3 \left[ \frac{1}{2r_a^2}(\partial_{t_a}\Phi)^2
-\frac{1}{2}\left(\partial_{r_a}\Phi\right)^2-U(\Phi)\right].
\end{align}

To be specific, we now consider the archetypical example for tunneling in field theory that is given by the quartic potential
\begin{align}
 \label{eq:Uarchetypical}U(\Phi)=U_0-\frac{1}{2}\mu^2\Phi^2+\frac{g}{3!}\Phi^3+\frac{\lambda}{4!}\Phi^4
\end{align}
 with $\mu, g, \lambda$ all taking positive real values and $g\rightarrow 0$ in the thin-wall regime~\cite{Coleman:1977py,Callan:1977pt}. The deformations of the Coleman bubble  can then be parametrized by making use of the kink profile
\begin{align}
\label{thin-wall}
\varphi_{\rm bubble}(r_a,t_a)\approx v\tanh[\gamma (r_a-R_a(t_a))],
\end{align}
where
\begin{align}
 v=&\,\sqrt{\frac{6\mu^2}{\lambda}}, & \gamma=&\,\frac{\mu}{\sqrt{2}}.
\end{align}
For the Coleman bubble, one has $R_a(t_a)=R_c$, with $R_c$ given by
\begin{align}\label{eq:RcColeman}
R_c=&\,\frac{12\gamma}{gv},
\end{align}
while the bounce action is $B(R_c)=8\pi^2 R_c^3\gamma^3/\lambda$~\cite{Garbrecht:2015oea,Ai:2018guc}.

Substituting Eq.~\eqref{thin-wall} into Eq.~\eqref{action-co} and taking the limit of large $R_a$ (consistent with the thin-wall approximation), we obtain
\begin{align}
\label{action-bubble-wall}
S_{\rm M}(R)&=4\pi\int\d t_a\,(\cosh^2 t_a)\left[\frac{2v^2\gamma {R_a}}{3}\left(\frac{\d R_a}{\d t_a}\right)^2\right.\notag\\
&\ \ \ \left. -\frac{1}{2\pi^2}B(R_a)\right],
\end{align}
where we have used Eq.~\eqref{bounce-action}. Since the radius $R_c$ of Eq.~\eqref{eq:RcColeman} is an extremum of $B(R_a)$, expanding the latter around $R_c$ gives
\begin{align}
\label{expan}
B(R_a)\approx B(R_c)+2B(R_c)\lambda_0(R_a-R_c)^2,
\end{align}
where we have used Eq.~\eqref{negative-eigenvalue}.
Substituting the above expansion into Eq.~\eqref{action-bubble-wall}, one obtains the equation of motion for the bubble wall
\begin{align}
\label{eom-acc}
\frac{\d^2 R_a}{\d t_a^2}+2(\tanh t_a)\frac{\d R_a}{\d t_a}+\frac{3B(R_c)\lambda_0}{2\pi^2v^2\gamma R_a}(R_a-R_c)=0.
\end{align}
In the above equation, we have neglected a term $(\d R_a/\d t_a)^2/2R_a$ which is suppressed since we are considering the large $R_a$ regime.
When one substitutes the expressions given below Eq.~\eqref{thin-wall} into the last term, one finds that the above equation of motion only depends on the parameter $R_c$,
\begin{align}
\label{eom-acc2}
\frac{\d^2 R_a}{\d t_a^2}+2(\tanh t_a)\left(\frac{\d R_a}{\d t_a}\right)-\frac{3R_c}{R_a}(R_a-R_c)=0.
\end{align}
Further taking $R_a=R_c+\delta R_a$, one obtains
\begin{align}
\label{eom-acc3}
\ddot{ \delta R_a}+2(\tanh t_a)\dot{\delta R_a}-3\delta R_a=0.
\end{align}
We clearly see that the negative mode gives rise to the last term above and drives the bubble wall away from $R_c$ (because $\lambda_0<0$).
Note that $R_a(t_a)\equiv R_c$ is a solution to Eq.~\eqref{eom-acc} with the initial condition $R_a(t_a=0)=R_c$ and $\d R_a/\d t_a|_{t_a=0}=0$. But this solution is unstable. Any small perturbation that gives the bubble wall a deviation from $R_c$ or from the vanishing velocity will trigger an increasing deviation of the bubble wall away from the position $R_a=R_c$.

The validity of equation of motion~\eqref{eom-acc} is limited by two assumptions that we have made. First, we consider here the thin-wall regime. Second, Eq.~\eqref{eom-acc} is valid only for $R_a\sim R_c$ since we have used the expansion~\eqref{expan}. Nevertheless, equation~\eqref{eom-acc} clearly shows that the instability of the motion for the bubble wall is induced by the negative eigenvalue $\lambda_0$ which remains present beyond the thin-wall regime.

In particular, for $\delta R_a<0$, $R_a$ will become smaller at an exponentially increasing rate. From the point of view of an accelerating observer, this would look like a contracting bubble. This, however, does not correspond to a failed nucleation from the point of view of an inertial observer. In the rest frame of nucleation, the bubble always expands.
On the other hand, for $\delta R_a>0$, the bubble moves away from criticality to values of $R_a>R_c$. The expanding and contracting behavior observed in the accelerating frame only represents two different possible deviations from the Coleman bubble. Given the standard arguments about the growth of tachyonic modes~\cite{Guth:1985ya}, the instability that is initially triggered through the quantum fluctuations about the parameter $R_c$ can be described after some time by a classical statistical ensemble of different trajectories $\delta R_a(t)$.

We note that it is the condition~$R_a(t_a=0)=R_c$ that breaks Lorentz invariance. The fluctuations are assumed to start growing at $t=0$, i.e. at the instant of bubble nucleation, where the wall is at rest. Then, the information about the rest frame of nucleation might remain contained in the growth of the instability.\footnote{If the bubble wall were subject to completely random perturbations during its history, then the information would certainly be lost in noise at late times.} A bubble moving at a relative velocity could then be described by shifting the parameter $t_a$ in the coordinate transformation~(\ref{frame:acc}). The growing fluctuation therefore determines the rest frame of nucleation in an observer-independent way. While we think that this interpretation of the growing instability is plausible, it would be interesting to follow up on the quantum-to-classical transition in the evolution of $R_a$ in more detail. For such a purpose, it may be useful to set up an effective Schrödinger equation for this parameter so that one can relate with the emerging classical statistical behavior for an upside-down harmonic oscillator~\cite{Guth:1985ya}.

There is a close analogy between false vacuum decay in the thin-wall regime and the Schwinger effect~\cite{Schwinger:1951nm,Sauter:1931zz,Heisenberg:1936nmg}. Both phenomena can be described as quantum tunneling of an effective relativistic particle~\cite{Ai:2020vhx}. In the thin-wall regime, thinking along the same lines, one can effectively describe the location of the bubble wall as a point particle, as we have done in order to set up the action Eq.~\eqref{action-bubble-wall}. As a check, we derive the equation of motion for the bubble wall in the accelerating frame using the point-particle like description. A similar use of the effective point-like description for the expansion of thin-wall bubbles at zero and finite temperature can be found in Refs.~\cite{Darme:2017wvu,Ellis:2019oqb,Cai:2020djd}.

The zero-temperature thin wall is governed by the following action (see, e.g., Ref.~\cite{Darme:2017wvu}),
\begin{align}
S_{\rm p}=\int\d t\left[-4\pi r^2\sigma\sqrt{1-\left(\frac{\d r}{\d t}\right)^2}+\frac{4\pi r^3}{3}\epsilon\right].
\end{align}
The kinetic term corresponds to the surface tension with the appropriate Lorentz factor and the potential term to the latent heat.
The radius $r$ appears here as a single degree of freedom and therefore behaves analogous to the trajectory of a relativistic point particle.
Performing the coordinate transformation~(\ref{frame:acc}), we obtain the Lagrangian
\begin{align}
L_{\rm p}=&-4\pi r_a^2\sigma\cosh^2t_a\sqrt{r_a^2-\dot{r}_a^2}\notag\\
&+\frac{4\pi r_a^3\epsilon}{3}\cosh^3t_a\left(\dot{r}_a\sinh t_a+r_a\cosh t_a\right),
\end{align}
where $\dot{r}_a$ denotes the derivative of $r_a$ with respect to $t_a$.
The corresponding Euler-Lagrange equation is
\begin{align}
\ddot{r}_a&-\frac{4\dot{r}_a^2}{r_a}+2(\tanh t_a)\dot{r}_a-2(\tanh t_a)\frac{\dot{r}_a^3}{r_a^2}\notag\\
&-\frac{3 r_a}{R_c}\sqrt{r_a^2-\dot{r}_a^2}+\frac{3r_a \dot{r}_a^2}{R_c r_a^2}\sqrt{r_a^2-\dot{r}_a^2}+3r_a=0.
\end{align}
Since $r_a\sim R_c\gg \dot r_a$, we have
\begin{align}
\ddot{r}_a+2(\tanh t_a)\dot{r}_a-\frac{3 r_a}{R_c}\sqrt{r_a^2-\dot{r}_a^2}+3r_a=0.
\end{align}
Expanding $r_a=R_c+\delta r_a$, we finally arrive at
\begin{align}
\ddot{\delta r}_a+2(\tanh t_a)\dot{\delta r}_a-3\delta r_a=0.
\end{align}
This is exactly the same as Eq.~\eqref{eom-acc3} as one would expect.

The instability of the bubble wall analytically continued from the Euclidean bounce solution is reminiscent of of the instability for the small black hole that appears in the Hawking-Page phase transition~\cite{Hawking:1982dh}. There, the small black hole is unstable because of its negative specific heat. For temperature ranges that allow for a gravitational first-order phase transition, there are typically three saddle points in the Euclidean path integral: two stable ones, namely the thermal anti-de Sitter (AdS) space and the large black hole in AdS, and an unstable one being the small black hole in AdS. The small black hole thus plays the role of the bounce in the Hawking-Page phase transition. It has one and only one negative mode in its fluctuation spectrum. When analytically continued to Lorentzian signature, the small black hole has negative specific heat and thus is unstable. This instability can then be seen as a consequence of the existence of the negative mode in the Euclidean formalism.

\subsection{Static bubble at finite temperature}

We now show the instability in the time evolution of perturbations of the static bounce during finite-temperature phase transitions.
In the analysis of zero temperature bubbles, we have seen that the Coleman bubble appears static in the uniformly accelerating frame, and the instability was shown to lead to the expansion or contraction of the perturbed bubbles.

In the case of the static bounce, as it is already time independent in the inertial plasma frame, one could carry out the same analysis as before by considering the time evolution of configurations with initial conditions close to the form of the static bounce in the plasma frame. However, in principle one should also consider the dynamics of the plasma. This requires for example to include the finite-temperature corrections to the potential $U(\Phi)$ mentioned before, as well as nonzero plasma velocities $v$. Then, one cannot model the bubble expansion through an equation of motion that is simply obtained from a Lagrangian for a scalar field with a potential that has constant coefficients. Nevertheless, close to the time of bubble nucleation, one may ignore variations of $T$ and $v$, so that one can consider the usual scalar dynamics with a potential $U(\Phi)$ with constant couplings, which may include finite-temperature corrections evaluated at the nucleation temperature. In this regime, we can perform an analysis analogous to the zero temperature case. To make contact with the discussion on the zero-temperature case, we still use Eq.~(\ref{eq:Uarchetypical}) to parametrize the potential.  Now consider a family of  deformations of the static bounce parametrized as $\varphi(r,t)=v \tanh[\gamma(r-R(t))]$ in the thin wall limit, with the profile given as in Eq.~\eqref{thin-wall}, with $r_a$ substituted by $r$, the same values of $v$ and $\gamma$,
\begin{align}\label{eq:Rcstatic}
	R_{c,\rm static}=\frac{8\gamma}{gv}.
\end{align}
 and $R_a(t_a)$ substituted with $R(t)=R_{c,\rm static}+\delta R(t)$. Substituting the ansatz into the three-dimensional action, and using Eq.~\eqref{eq:Rcstatic} one obtains the following equation of motion for $\delta R$:
\begin{align}
\delta \ddot{ R}-\frac{2}{R_c^2} \,\delta R=0.
\end{align}
As it occurs for perturbations of the Coleman bubble in the uniformly accelerating frame, there is an unstable behavior by which  supercritical bubbles ($\delta R>0$) expand, and subcritical bubbles ($\delta R<0$) collapse.

\begin{figure}

  \centering

  \includegraphics[scale=0.5]{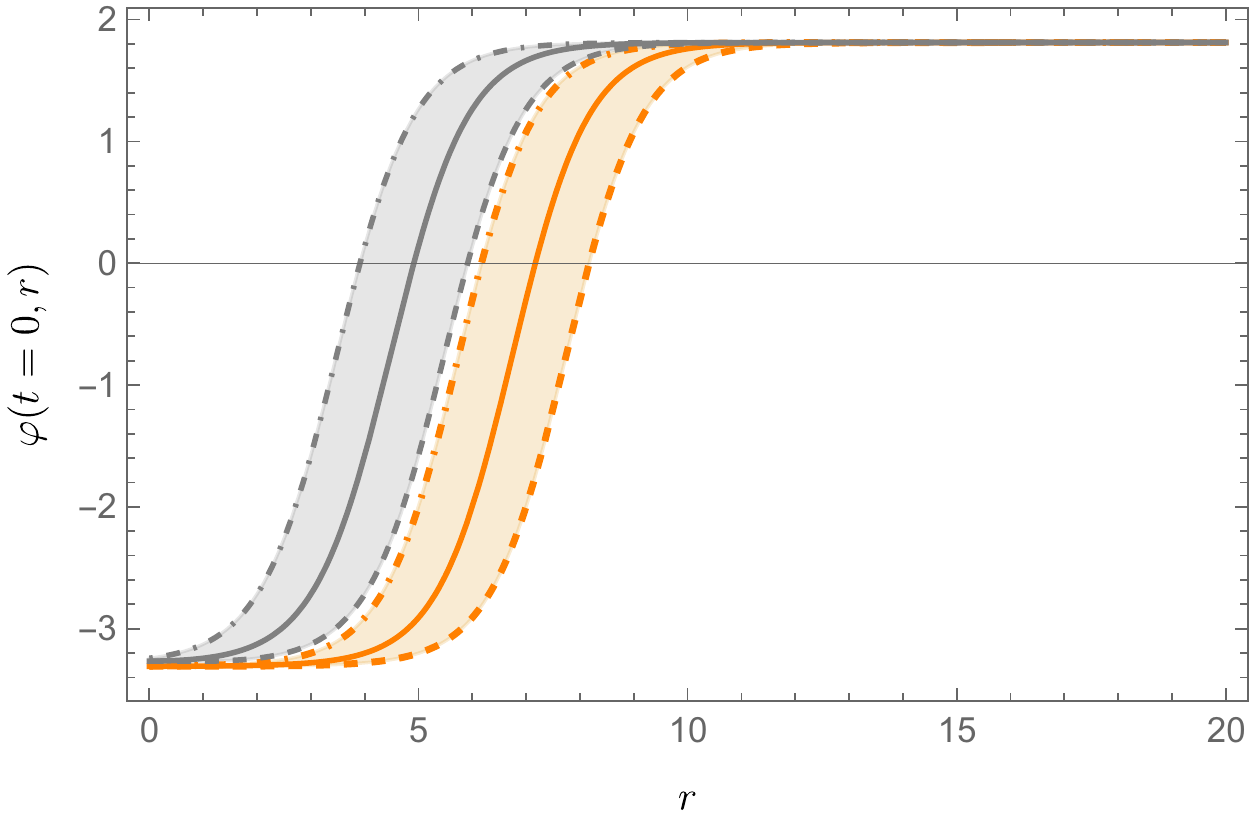}

  \caption{Orange: continuation of the Coleman bounce in the numerical example for $t=0$ (solid line), plus deformations with larger (dashed line) and smaller (dot-dashed line) bubble radius. Gray: analogous curves for the static bounce. \label{fig:bounces}}

\end{figure}

\begin{figure}

  \centering

  \includegraphics[scale=0.5]{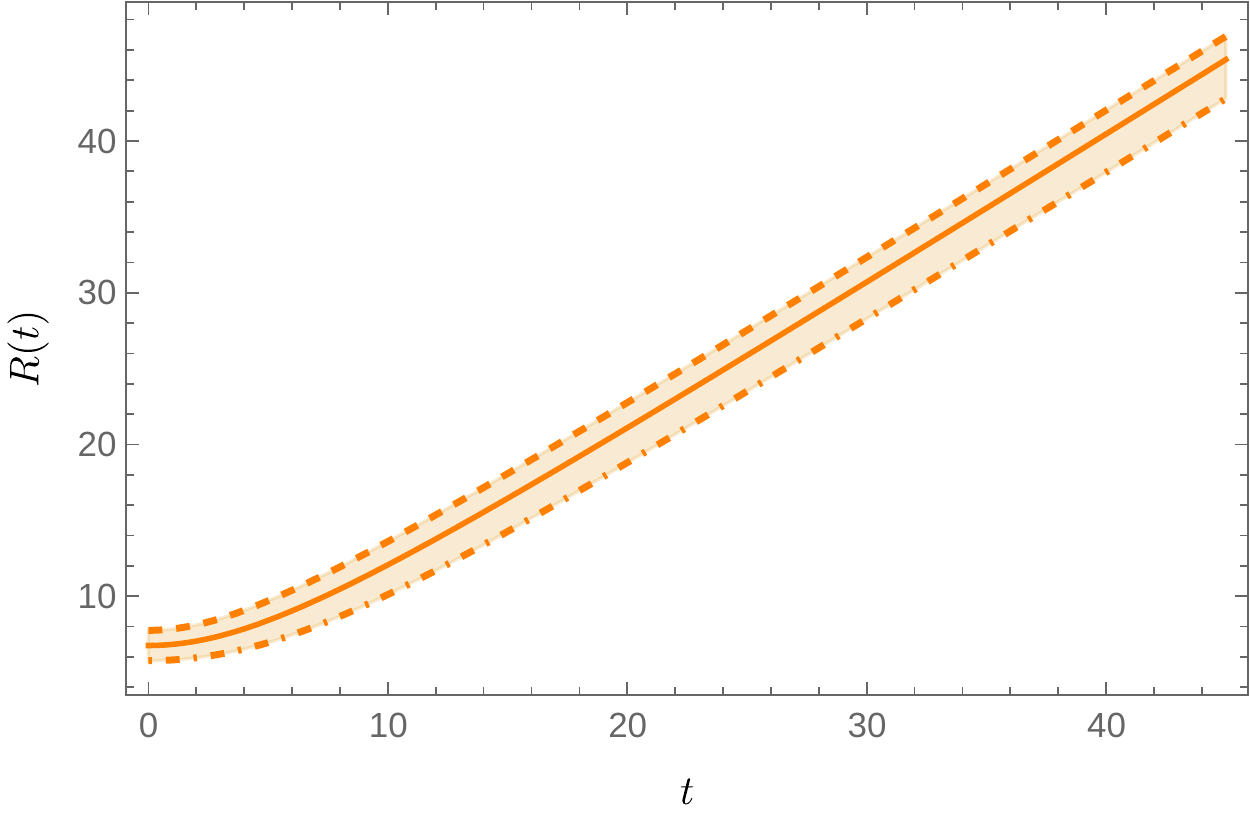}
   \includegraphics[scale=0.5]{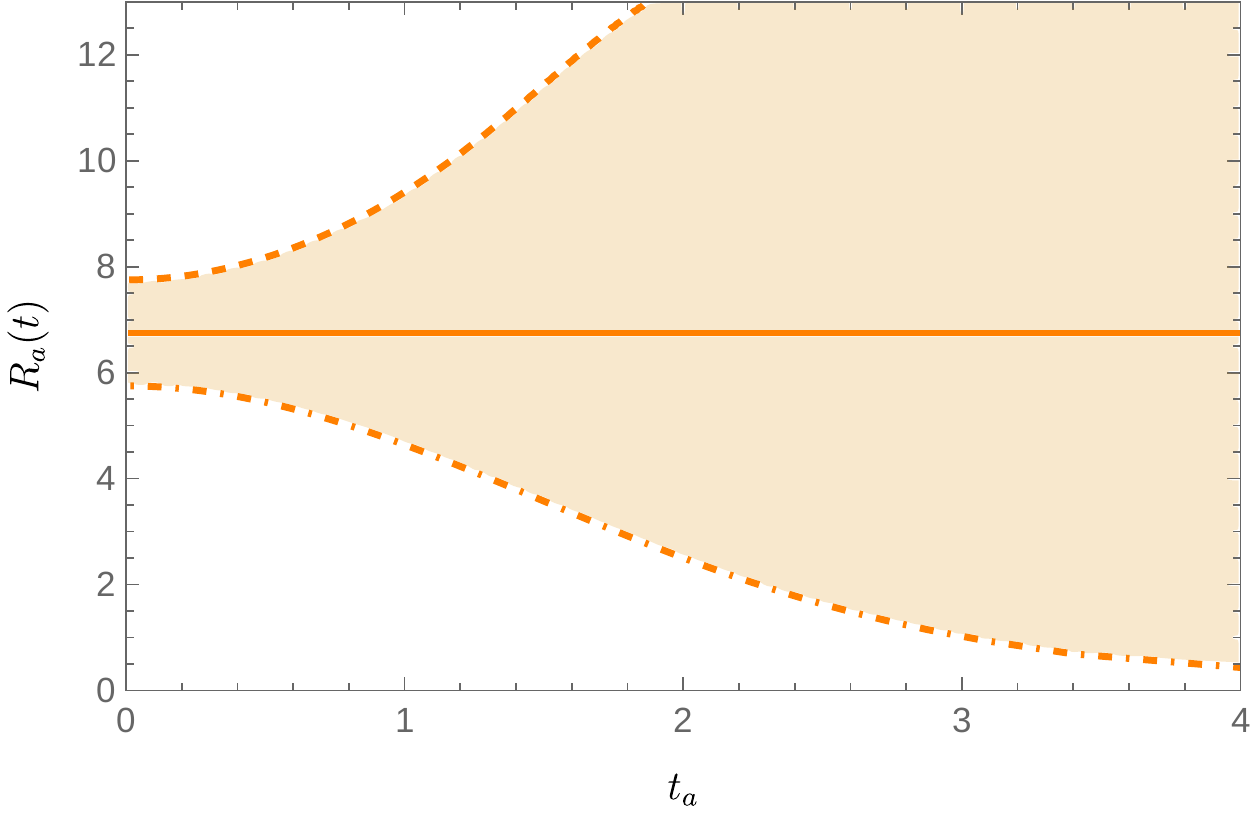}
   \includegraphics[scale=0.5]{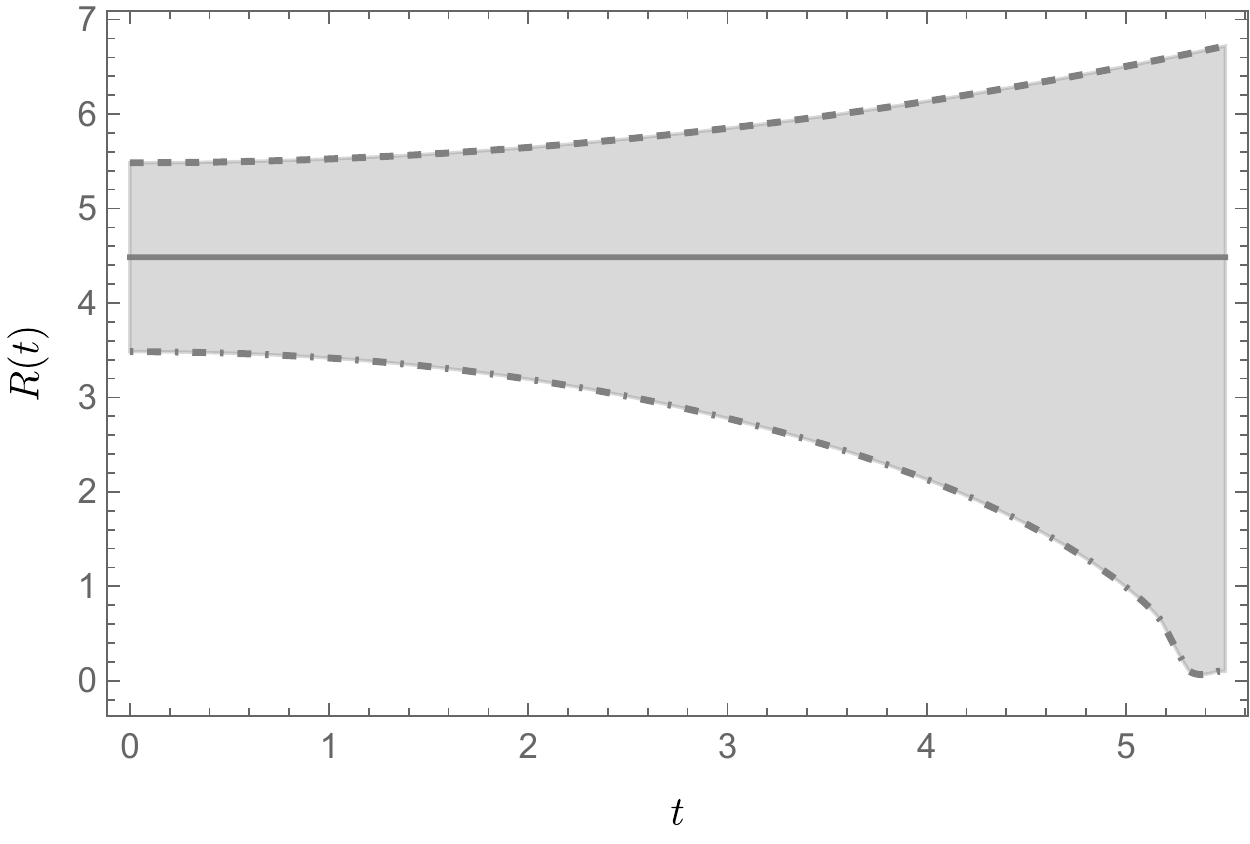}

  \caption{Evolution of the radii for the bubbles with initial conditions as in Fig.~\ref{fig:bounces}. The upper and middle plots give the evolution of the deformations of the Coleman bubble in the rest frame of nucleation and accelerating frame, respectively. The bottom plot gives the evolution of the deformations of the static bubble in the plasma frame. \label{fig:radii}}

\end{figure}

To close this section, let us provide a numerical confirmation, which does  not rely on the thin wall regime, of the instability of bubble propagation at zero and finite temperature by studying the time evolution of bubble profiles obtained by deforming the Coleman and static bubbles at $t=0$. For this purpose, we choose the parameters in Eq.~\eqref{eq:Uarchetypical} as
$$\mu=1,\quad g=1/2\quad \lambda=1,$$
and
\begin{equation*}
	U_0=-\frac{1}{256} (-537 + 35 \sqrt{105})=0.696706.
\end{equation*}
This leads to $U(\varphi)$ having a shape in qualitative agreement with Fig.~\ref{fig:potential}, with
\begin{align*}
	\varphi_+&=\frac{1}{4} \left(\sqrt{105}-3\right)=1.81174,\\
	\varphi_-&=-\frac{1}{4} \left(\sqrt{105}+3\right)=-3.31174.
\end{align*}
The Coleman bounce continued to Minkowski spacetime at $t=0$, $\varphi_{\rm bubble}(\rho=\sqrt{-t^2+r^2})|_{t=0}$, is shown by the orange line in Fig.~\ref{fig:bounces} together with two deformations in which the bubble radius is altered by one unit. We choose to define the bubble radius $R(t)$ as the value of $r$ for which $\varphi_{\rm bubble}(r,t)$ reaches $1/2(\varphi(0,0)+\varphi(\infty,0)$). The figure also shows the static bounce in gray, plus two deformations in which the radius is modified by $\pm 1$. Note how, when the same potential is used for the zero-temperature and finite-temperature phase transitions, the radius of the static bubble is smaller than that of the Coleman bubble, as it is also indicated by the thin-wall results of Eqs.~\eqref{eq:RcColeman} and \eqref{eq:Rcstatic}. In fact, the previous equations match the numerical values for the radii up to 3\% deviations. Taking the profiles in Fig.~\ref{fig:bounces} as initial conditions, together with the requirement $\dot{\varphi}(t=0,r)=0$, we can solve the  evolution equations for the scalar field in the  rest frame of nucleation (at zero temperature) and plasma frame (at finite temperature) under the assumption of an $O(3)$ symmetry,
\begin{align}
 \ddot{\varphi}(r,t)-\frac{\d^2}{\d r^2}\varphi(r,t)-\frac{2}{r}\frac{\d}{\d r} \varphi(r,t)+U'(\varphi)=0.
\end{align}
Doing so  we obtain the bubble radii illustrated in Fig.~\ref{fig:radii}.
The top and middle plots show the evolution of the deformed Coleman bubbles in the rest frame of nucleation and the accelerating frame, respectively. The bottom plot shows the evolution of the deformations of the static bubble in the plasma frame. The results confirm the expected unstable behavior, in which the deformations of the bubble configurations become amplified in the cases where the initial one was static. In the case of the Coleman bubble, note that the unstable behavior does not appear so dramatic when seen from the rest frame of nucleation, in which all the deformed bubbles considered here keep expanding. However in the accelerating frame, in which the Coleman bubble is static, only the supercritical bubbles expand, while the subcritical ones contract. A similar behavior is observed for deformations of the static bubble in the plasma frame.  Note that, were we to consider larger deformations of the Coleman bubble, one would eventually find bubbles that collapse. This is because such deformations would eventually fall into the subcritical region of the static bounce, as is clear from Fig.~\ref{fig:bounces}. Our results suggest a correspondence between the dynamics of Coleman bubbles in the uniformly accelerating frame, and finite-temperature phase transitions: both cases involve static H-critical bubbles, and feature an unstable behavior under perturbations. In the next section, we will elaborate further on this correspondence.

\section{Correspondence between false vacuum decay at zero and finite temperature}

\label{sec:correspondence}

In this section, we explain the relation between the static vacuum bubble configuration in the accelerating frame and the static bubble configuration in a finite-temperature plasma. This relation also leads to an interpretation of the negative modes about the Coleman bounce in terms of the instability of the finite-temperature bubble.

\begin{figure}
\centering

\includegraphics[scale=0.5]{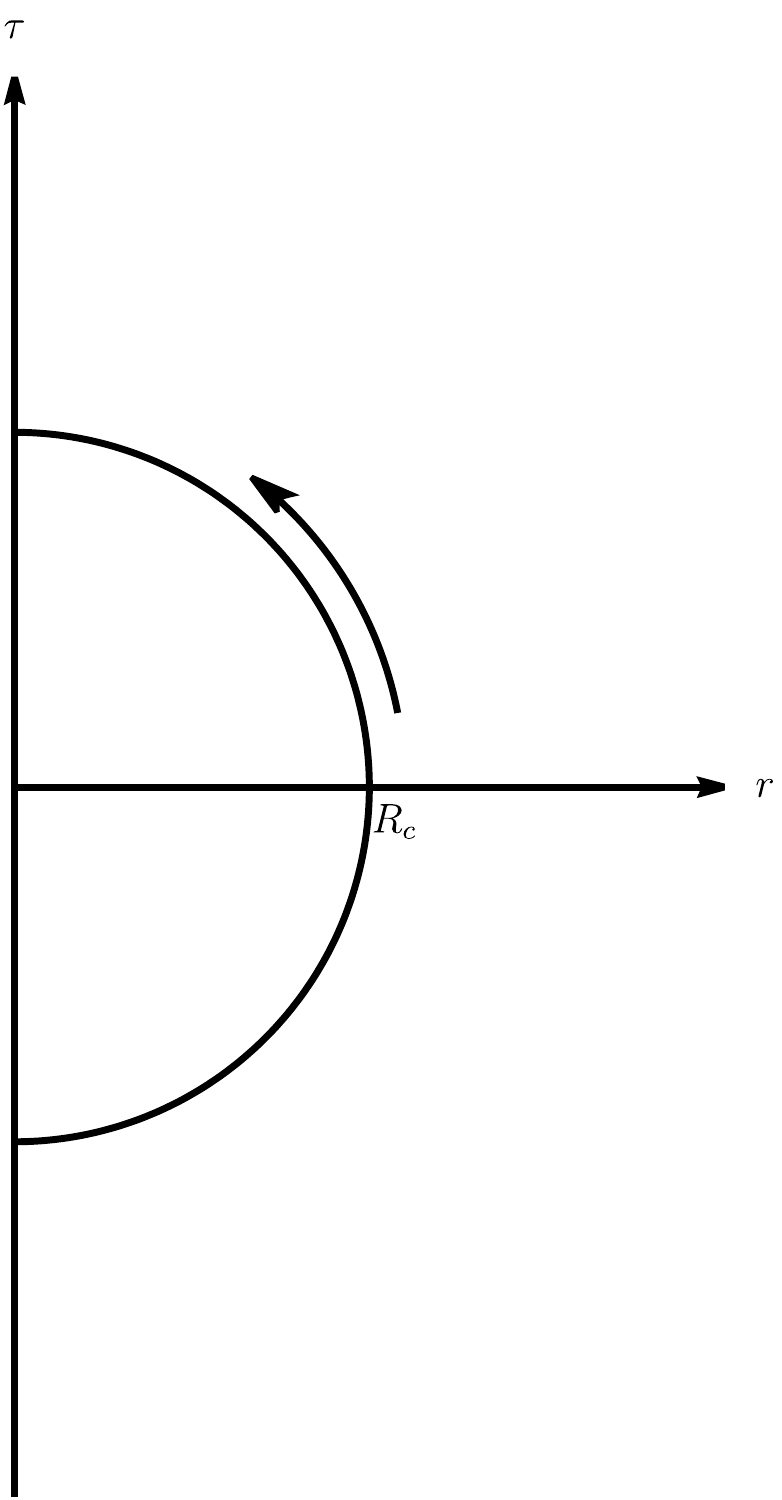}
\caption{The $O(4)$-symmetric bounce can be viewed as a static bounce with Euclidean time $\tau_a$. The solid half circle represents the phase boundary between the false vacuum (outside) and the true vacuum (inside).
\label{fig:staticbounce}}
\end{figure}

If one performs a Wick rotation of the time parameter $t_a\rightarrow -{\rm i}\tau_a$, then the original $O(4)$ bounce becomes static with respect to the Euclidean time $\tau_a$, as shown in Fig.~\ref{fig:staticbounce}.  As was mentioned previously, this is reminiscent of the static bounce for false vacuum decay at finite temperature. Therefore, one  may think that bubble nucleation seen by the accelerating observers is a thermal vacuum transition. In that case, the same process has two different descriptions: in the rest frame of nucleation and the accelerating frame. The equations of motion must coincide in order to give the same bounce. It should be noted that when staying within the rest frame of nucleation, the critical configurations corresponding to the Coleman bounce and the static bounce associated with thermal transitions are inequivalent and solve different equations, i.e., Eqs.~~\eqref{Eq7} and~\eqref{eq:staticbubble}, respectively. The fact that the static bounce in the accelerating frame satisfies the same equation as the Coleman bounce is made possible by the nontrivial spacetime metric in the noninertial frame. The duality between vacuum and thermal transitions has already been pointed out in Ref.~\cite{Ai:2018rnh} when the bubble is nucleated around a horizon.\footnote{For black hole horizons, this correspondence is valid only for the Hartle-Hawking vacuum~\cite{Shkerin:2021zbf}.} The existence of a horizon provides a thermal description for the static observers outside of the horizon (which are uniformly accelerating observers for the Rindler horizon). However, the situation here is more subtle.

\begin{figure}
\centering
\includegraphics[scale=0.45]{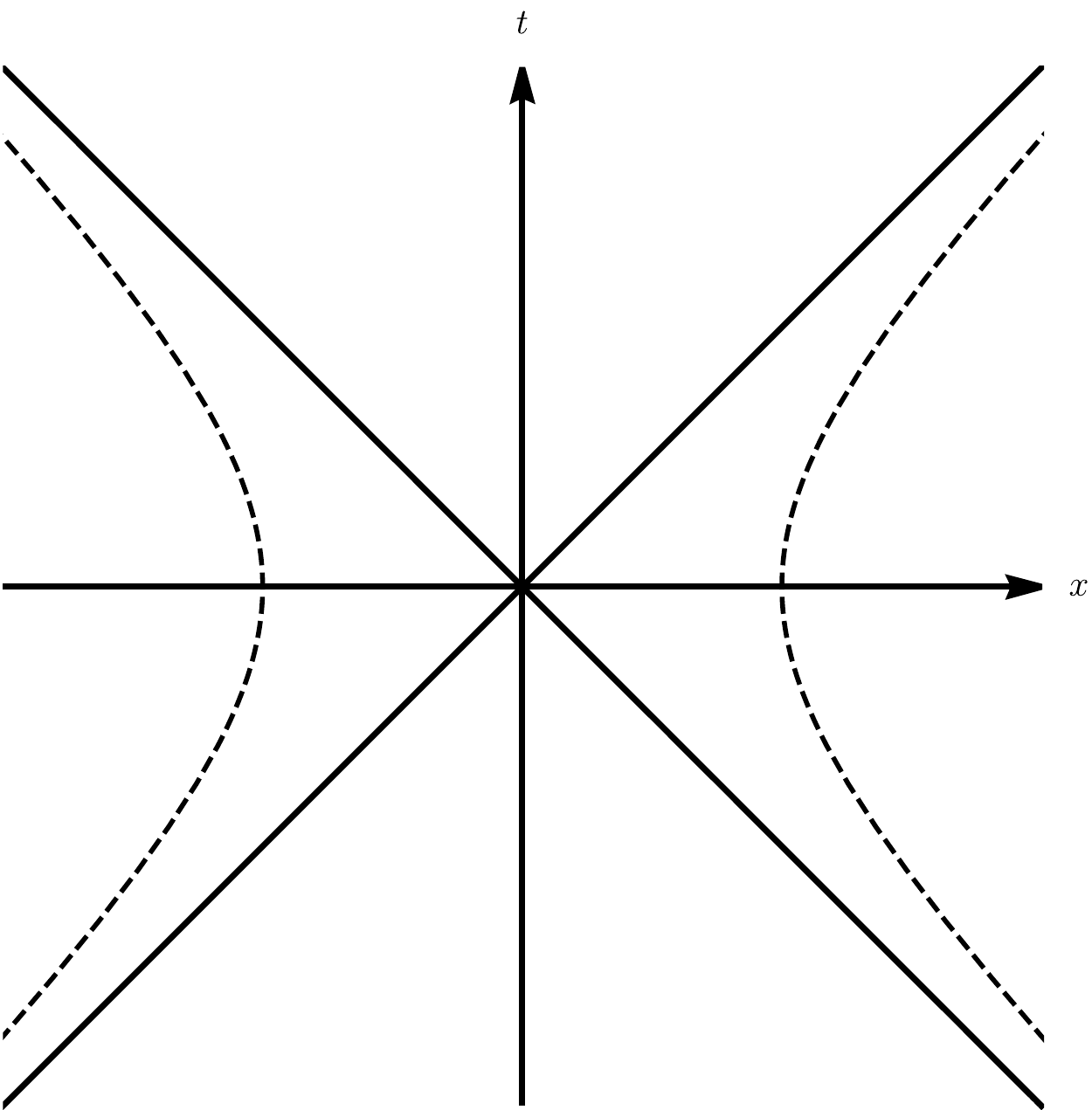}
\caption{Rindler horizon for the uniformly accelerated observers in the $x$ direction. The $y$ and $z$ directions are suppressed. The horizon is invariant under the $y$ and $z$ translations.
\label{fig:Rindler}}
\end{figure}

Given the form of the Coleman bounce,
the Euclidean time $\tau_a={\rm i}t_a$,
does not allow for a symmetry of  the bounce  background in Euclidean spacetime of the form $O(2)\times G$, with $O(2)$ acting in the imaginary time direction and $G$ on the spatial coordinates. Thus, one cannot identify a globally thermal system. This is consistent with the fact that there is no horizon in the frame $\{t_a,r_a,\theta,\phi\}$. Note that the light cone $r=t$ (or $r_a=0$) is not a Rindler horizon. To have a Rindler horizon, one may consider a Rindler frame $\{t_R,x_R,y,z\}$ obtained from
\begin{subequations}
\begin{align}
t= x_R \sinh t_R,\\
x= x_R\cosh t_R,
\end{align}
\end{subequations}
where $\{t,x,y,z\}$ are the inertial coordinates from the rest frame of nucleation.
The Rindler horizon is shown in Fig.~\ref{fig:Rindler}. The essential difference between a Rindler horizon and the light cone $r=t$ is that a Rindler horizon separates two causally uncorrelated regions, the wedges $(x>0,-x<t<x)$ and $(x<0, -x<t<x)$. Consider the spatial slice at $t=0$. The state on $x>0$ is entangled with the state on $x<0$. For the Rindler observers in one single wedge, another wedge is unobservable and must be traced over. The reduced density matrix is thermal~\cite{Unruh:1976db}. In our case, the accelerating frame with spatial spherical symmetry fills all of the space at $t=0$. A pure state defined on this spatial hypersurface therefore remains pure when restricted to the frame~(\ref{frame:acc}). This is why we do not have a thermal interpretation for the vacuum transition that is globally valid in the radially accelerating frame.

However, any {\it single} accelerating observer with given $r_a,\theta,\phi$ in the radially accelerating frame~(\ref{frame:acc}) is indistinguishable from a Rindler observer. Therefore any local consequences of the Unruh effect must also be experienced by observers that are static in the accelerating frame $\{t_a,r_a,\theta,\phi\}$. Such observers see a {\it locally} thermal quantum field system. Pictorially, the local thermal property can be thought of as originating from the entanglement between the antipodal points, $\{r_a,\theta,\phi\}$ and $\{r_a,\pi-\theta,\phi+\pi\}$. In this sense, we may extend the correspondence proposed in Ref.~\cite{Ai:2018rnh} to the situation analyzed in the present paper. Namely, the bubble nucleation at zero temperature observed in the rest frame of nucleation has {\it locally} a thermal description in the accelerating frame $\{t_a,r_a,\theta,\phi\}$. Since the static bounce from the thermal description and the $O(4)$-symmetric bounce pertain to the same transition process, the equations of motion must coincide. (The mismatch between the global symmetries in both interpretations does not affect the requirement on the coincidence of the equation of motion.) The profile of the static bubble is simply obtained from the analytic continuation of the static bounce solution in the thermal description. Because of the stationary condition, the time variables do not appear in the equation of motion. This explains why Eq.~\eqref{Eq25} takes the same form as Eq.~\eqref{Eq7}.

From the theory of finite-temperature phase transitions we know that there must be a negative mode for the fluctuations about the static bounce which gives an imaginary part for the free energy. This negative mode is inherited when we perform the inverse Wick rotation $\tau_a\rightarrow {\rm i} t_a$ again, because the time coordinates do not appear in the eigenvalue equations for the fluctuation spectrum. The uniform acceleration of the bubble is then unstable because of the negative mode. Given that the existence of a negative mode for the fluctuations about the static bounce is a necessary condition for false vacuum decay, the instability in the uniform acceleration is robust.

\section{Conclusions}
\label{sec:conclusion}

False vacuum decay plays an important role in a variety of phenomenological studies. In particular, vacuum transitions in the early Universe could be a source of gravitational waves by means of collisions and mergers of the nucleated bubbles and therefore currently evoke an increasing interest in the cosmology community. For most studies, the bubble motion is particularly relevant. In this paper, we revisit the bubble growth for false vacuum decay at zero  temperature and at finite temperature shortly after nucleation. In the case of zero temperature, we have shown that the picture of uniformly accelerating bubble expansion from Coleman needs to be supplemented. We observe that the standard uniformly accelerated bubble, although satisfying the classical equation of motion, is not stable under perturbations preserving the spherical symmetry. This is demonstrated by studying the eigenvalue equation for fluctuations in the static background of a Coleman bubble as seen from a uniformly accelerating frame and prove the existence of a tachyonic mode in the spectrum. In this accelerating frame, in which the critical Coleman bubble is static, the instability manifests itself in the growth (decrease) of  bubbles having greater (smaller) radii than the radius given by the S-critical bubble. The instability is related to the well-known negative mode of the fluctuation operator of the Euclidean action about the Coleman bounce, which plays a fundamental role in the computation of the tunneling rate.~Earlier studies of instabilities focused on deformations violating the spherical symmetry, or considered configurations with multiple bubbles~\cite{Adams:1989su, Garriga:1991ts, Garriga:1991tb, Bond:2015zfa,Bond:2015zfa}.

At finite temperature, we have shown that as long as the changes in the plasma velocity and temperature are small, as expected in the early stages of bubble growth, there is an analogous instability in the plasma frame, by which perturbations of the critical static bubble with larger or smaller radii respectively expand or collapse. This instability is of course well known and is the reason why the static bubble is called critical (see Sec.~\ref{sec:standard} for the terminology). However, a less emphasized point, sometimes being misunderstood, is that the Coleman bubble in the rest frame of nucleation is not critical but only S-critical. Only in the accelerating frame it can be interpreted as H-critical. The unstable behavior for the Coleman bubble in the accelerating frame is completely analogous to that of static bubbles in the plasma frame. This indicates that vacuum transitions viewed from accelerating frames can be seen as finite-temperature phase transitions~\cite{Ai:2018rnh}. This duality suggests that, in the same way that simulations of finite-temperature phase transitions use bubble profiles different from the exact H-critical bubble to ensure the growth after nucleation, when simulating vacuum transitions one may also consider bubble configurations differing from the exact S-critical Coleman bubble. In this case, the instability studied in this paper would manifest itself in the simulations, although in the rest frame of nucleation this effect may be very small.

\begin{table*}[ht]
    \begin{tabular}{|c|c|c|c|}
    \hline
         & Zero T, rest frame of nucleation   & Zero T, comoving accelerating frame & Finite T, plasma frame \\
    \hline         
   Nucleated bubble & S-critical & H-critical & H-critical \\
    \hline
    Thermal bath & No & Unruh bath &  Physical thermal plasma\\
    \hline
   Global ``temperature'' & 0 & $1/2\pi$ & $T$ of the plasma \\
    \hline
   Motion without perturbations & Uniformly accelerating & Static & Static \\
    \hline
   Instability & Of the uniform acceleration & Of the staticity & Of the staticity \\
    \hline
    \end{tabular}
    \caption{The dual picture of zero-temperature phase transitions in terms of finite-temperature phase transitions and the instability discussed in this work.}
    \label{tab:label1}
\end{table*}

Note that under the duality mentioned above, the instability we discovered for the uniformly accelerating bubble at zero temperature is mapped to the instability of the nucleated critical bubble at finite temperature, instead of the instability in its latter growth.~We summarise the dual relations in Table~\ref{tab:label1}. In the case of finite-temperature phase transitions, one can have additional instabilities in the temperature and velocity fields of the plasma~\cite{Huet:1992ex} in the bubble growth.

Finally, we note that under the plausible assumption that the instability starts to develop at the instant of nucleation, the growing fluctuations carry the information about the rest frame of nucleation and are important for understanding the development of the bubble away from criticality in an observer-independent way.

\section{Acknowledgments}

W.Y.A. is supported by the UK Engineering
and Physical Sciences Research Council under Research Grant No. EP/V002821/1. C.T. acknowledges financial support by the DFG through the ORIGINS cluster of excellence. The work of J.S.C. was supported by the Research Fund Denmark Grant No. 0135-00378B.





\bibliography{FVDref}

\end{document}